\documentclass[letter,oldversion]{aa}
\usepackage{natbib}
\bibpunct{(}{)}{;}{a}{}{,}
\usepackage{graphicx}
\usepackage{revsymb}	
\usepackage{amsmath}	
\usepackage[usenames]{color}	
\usepackage{epstopdf}	
\newcommand{\eq}[1] {Eq.\,(\ref{#1})}

\usepackage{txfonts}
\begin{document}
       \title{Turbulent eddy-time-correlation in the solar convective zone}
         
\author{K. Belkacem\inst{1,2} \and R. Samadi\inst{2} \and M.J. Goupil\inst{2} \and F. Baudin\inst{3} \and D. Salabert\inst{4,5} \and T. Appourchaux\inst{3}}

\institute{Institut d'Astrophysique et de G\'eophysique, Universit\'e de Li\`ege, All\'ee du 6 Ao\^ut 17-B 4000 Li\`ege, Belgium.
\and 
 LESIA, UMR8109, Universit\'e Pierre et Marie Curie, Universit\'e
	Denis Diderot, Obs. de Paris, 92195 Meudon Cedex, France.
\and 
Institut  d'Astrophysique Spatiale, CNRS-Universit\'e Paris XI UMR8617, 91405 Orsay 
Cedex, France.
\and 
Instituto de Astrof\'isica de Canarias, E-38200 LaLaguna, Tenerife, Spain.
\and
Departamento de Astrof\'isica, Universidad de La Laguna, E-38206 La Laguna, Tenerife, Spain.
}
   
   \offprints{K. Belkacem}
   \mail{Kevin.Belkacem@ulg.ac.be}
   \date{}

  \abstract
  {
Theoretical modeling of the driving processes of solar-like oscillations is a powerful way of  understanding the properties of the convective zones of solar-type stars. In this framework, the description of the temporal correlation between turbulent eddies is an essential ingredient to model mode amplitudes. However, there is a debate between a Gaussian or Lorentzian description of the eddy-time correlation function (Samadi et al. 2003, Chaplin et al. 2005). 
Indeed, a Gaussian description reproduces the low-frequency shape of the mode amplitude for the Sun, but is unsatisfactory from a theoretical point of view (Houdek, 2009) and leads to other disagreements with observations (Samadi et al., 2007). These are solved by using a Lorentzian description, but there the low-frequency shape of the solar observations is not correctly reproduced.  
 We reconcile the two descriptions by adopting the sweeping approximation, which consists in assuming that the eddy-time-correlation function is dominated by the advection of eddies, in the inertial range, by energy-bearing eddies.  
Using a Lorentzian function together with a cut-off frequency derived from the sweeping assumption  allows us to reproduce the low-frequency shape of the observations. This result also constitutes a validation of the sweeping assumption for highly turbulent flows as in the solar case. 
}


   \maketitle
%

\section{Introduction}
\label{intro}

Excitation of solar-like oscillations is attributed to turbulent
motions that excite p modes \citep[for a recent review, see][]{Samadi09}. 
Their amplitudes  
result from a balance between excitation and damping 
and crucially depend on the way the eddies are temporally correlated 
as shown for solar $p$ and $g$ modes \citep{Samadi02I,Belkacem09,Appourchaux10}, for main-sequence stars \citep{Samadi10a,Samadi10b}, for red giants \citep{Dupret09}, or for massive stars \citep{Belkacem09b,Belkacem10b}. Hence, the improvement of our understanding and modeling of the temporal correlation of turbulent eddies, hereafter denoted in the Fourier domain as $\chi_k(\omega)$, is fundamental to infer turbulent properties in stellar convection zones. 

There are two ways to compute the eddy-time correlation function. A direct computation from 3D numerical simulations is possible and was performed by \cite{Samadi08}. 
Nevertheless, \cite{Samadi09} pointed out that the results depend on the spatial resolution, and therefore dedicated high-resolution 3D numerical simulations are required. This then becomes an important limitation when computing mode amplitudes for a large number of stars, preventing us from applying statistical astereosismology. 

The second way to compute $\chi_k$ consists in using appropriate analytical descriptions. 
Most of the theoretical formulations of mode excitation explicitly or implicitly assume
a Gaussian functional form for $\chi_k(\omega)$ \citep[][]{GK77,Dolginov84,GK94,B92c,Samadi00II,Chaplin05}. 
However, 3D hydrodynamical simulations of the outer layers
of the Sun show that at the length-scales close to those of the energy-bearing 
eddies (around $1$ Mm), $\chi_k$ is a Lorentzian function  \citep{Samadi02II,Belkacem09}.  
As pointed out by \citet{Chaplin05}, a Lorentzian  $\chi_k$  is also a
result predicted for the largest, most-energetic eddies by the
time-dependent mixing-length formulation derived by \citet{Gough77}. 
Therefore, there is numerical, theoretical, and also observational evidence  
\citep{Samadi07} that   $\chi_k$
is Lorentzian. 

However, \cite{Chaplin05}, \cite{Samadi09}, and \cite{Houdek09} found that 
 a Lorentzian $\chi_k$, when used with a mixing-length description of the whole convection zone, results in a severe over-estimate for the low-frequency modes. 
In this case, low-frequency modes ($\nu < 2 $mHz) are excited deep in the solar convective region by large-scale eddies that give a substantial fraction of the energy injected to the modes. 
\cite{Chaplin05} and \cite{Samadi09} then suggested that most contributing eddies situated deep in the Sun have a $\chi_k$ more Gaussian than Lorentzian because at a fixed frequency, a Gaussian $\chi_k$ decreases more rapidly with depth. 
 
We therefore propose a refined description of the eddy-time correlation function based on the sweeping approximation to overcome this issue.  
This consists in assuming that the temporal correlation of the eddies, in the inertial subrange, is dominated by the advection by energy-bearing eddies.  This assumption was first proposed by \cite{Tennekes75}, and was subsequently investigated by \cite{Kaneda93} and \cite{Kaneda99}. In this letter, we demonstrate that the low-frequency shape of the observed energy injection rates into the solar modes is very sensitive to this assumption and more precisely to the Eulerian microscale, defined as the curvature of the time-correlation function at the origin. Hence, modeling of the solar p-mode amplitudes is shown to constitute an efficient test for temporal properties in highly turbulent flows.   

The paper is organized as follows: Section~\ref{chiw} defines the eddy-time correlation function.  
In Sect.~\ref{sweeping}, we propose a short-time expansion of the eddy-time correlation function. 
The use of the Eulerian microscale as a cut-off is introduced in the computation of solar $p$ mode amplitudes and the result is compared to the observations in Sect.~\ref{amplitude}. Finally, Sect.~\ref{conclu} is dedicated to conclusions and discussions. 

\section{The  Eulerian eddy time-correlation function} 
\label{chiw}

For a turbulent fluid, one defines the Eulerian eddy time-correlation function as
\begin{align}
 \left< \vec u(\vec x + \vec r,t+\tau) \cdot \vec u(\vec x,t) \right> = \int  {\cal E}(\vec k,t,\tau) \, e^{{\rm i} \vec k \cdot \vec x} \, {\rm d}^3\vec k  \, ,
\label{time-correlation}
\end{align}
where $\vec u$ is the Eulerian turbulent velocity field, $\vec x$ and $t$ the space and time position of the fluid element, $\vec k$ the wave number vector, $\tau$ the time-correlation length, and $\vec r$ the space-correlation length. 
The function $\cal E$ in the RHS of \eq{time-correlation} represents the  time-correlation function associated with an eddy of wave-number $\vec k$.

We assume an isotropic and stationary turbulence, accordingly  ${\cal E}$ is only a function of $k$ and $\tau$.
The quantity ${\cal E}(k,\tau)$ is related to the turbulent energy spectrum according to
\begin{align}
{\cal E}(k,\tau) = \frac{E(k,\tau)}{2\pi k^2} \, ,
\end{align}
where $E(k,\tau)$ is the turbulent kinetic energy spectrum whose temporal Fourier transform is 
\begin{align}
\label{Ekw}
E(k,\omega)   \equiv {1 \over {2\pi}} \,  \int_{-\infty}^{+\infty}  {E}(k,\tau) \, e^{{\rm i} \omega \tau} \, {\rm d} \tau \, ,
\end{align}
where $\omega$ is the eddy frequency, and $E(k,\omega)$ is written following an approximated form proposed by \cite{Stein67}
\begin{align}
\label{decomp_E}
E(k,\omega) = E(k) \, \chi_k (\omega) \quad {\rm with}\quad  \int_{-\infty}^{+\infty} \chi_k (\omega) \, {\rm d} \omega = 1 \, ,
\end{align}
where $\chi_k(\omega)$ is the frequency component of $E(k,\omega)$. In other
words, $\chi_k(\omega) $  represents -~  in the frequency  domain ~-  the temporal correlation between eddies of wave-number $k$. 

As already discussed in Sect.~\ref{intro}, theoretical and observational evidence show that $\chi_k(\omega)$ is Lorentzian, \emph{i.e.} 
\begin{align}
\chi_k (\omega)  = { 1 \over {\pi \omega_k} } \, {1 \over {1 + \left ( \omega/\omega_k \right )^2} } \, ,
\label{lorentzian}
\end{align}
where $\omega_k$ is by definition the width at half maximum of $\chi_k(\omega)$. In the framework of \citet{Samadi00I}'s formalism, this latter quantity is evaluated as:
\begin{align}
\label{omega_k}
\omega_k = k \, u_k \quad {\rm with}\quad u_k^2 = \int_{k}^{2k} E(k) \, {\rm d}k \, ,
\end{align} 
where $E(k)$ is defined by \eq{decomp_E}. 
However, in the high-frequency regime ($\omega \gg \omega_k$), corresponding to the short-time correlation ($\tau \approx 0$), the situation is less clear. We next investigate short-time  correlations ($\tau \approx 0$).

\section{The sweeping assumption for the Eulerian time-correlation function} 
\label{sweeping}

\subsection{Short-time expansion of the eddy-time correlation function}
\label{high_freq}

The function ${\cal E}(k,t,\tau)$ (\eq{time-correlation}) can be expanded for short-time scales, in the inertial sub-range (\emph{i.e} for $k > k_0$ and $k < k_d$, where $k_0$ is the wave number of energy-bearing eddies and $k_d$ is the wave-number of viscous dissipation), using the Navier-Stokes equations and the sweeping assumption, as \citep[see ][ for a derivation]{Kaneda93}
\begin{align}
\label{DL}
{\cal E}(k,\tau) = {\cal E} (k,\tau=0) \, \left ( 1  -  \alpha_k  {\left | \tau  \right |}  -  \frac{1}{2} \,  \left ( { \omega_E \tau } \right )^2  + \dots \right ) \, , 
\end{align}
where the  characteristic frequency $\alpha_k$ is defined by the relation
\begin{align}
\epsilon = - \frac{1}{2} \frac{{\rm d}}{{\rm d} t}\left< \vec u \cdot \vec u \right> = \int   \alpha_k \, {\cal E} (\vec k,\tau=0) \, {\rm d}^3 \vec k 
\end{align}
with $\epsilon$ the dissipation rate of energy. Hence, $\alpha_k$ is the typical frequency of energy dissipation at the scale $k$. It can be estimated by assuming that a large fraction of the kinetic energy of eddies is lost within one turnover time \citep{Tennekes_Lumley72}. 
Hence, $\alpha_k$ is approximated by the turn-over frequency $\alpha_k= k \, u_k =\omega_k$ (see \eq{omega_k}). 

The second characteristic frequency, $\omega_E (k)$, is the curvature of the correlation function near the origin \citep[][]{Kaneda93}, and is defined by 
\begin{align}
\label{tau_E}
\omega_E = k \, u_0 \, .
\end{align}
The  associated characteristic time $\tau_E (k)= 2\pi \, \omega_E^{-1}$ is also referred to as the Eulerian micro-scale\footnote{It is the time equivalent of the Taylor micro-scale, which  corresponds to the largest scale at which viscosity affects the dynamic of eddies.} \citep{Tennekes_Lumley72}. 
An approximate expression for  $\omega_E (k)$ can be obtained by assuming the random sweeping effect of large eddies on small eddies.
This assumption consists in assuming that the velocity field $\vec{u}(k)$ associated with an eddy of  wave-number $\vec k$ lying in the inertial-subrange (\emph{i.e.} large $\vec k$ compared to $\vec k_0$) is advected by  the energy-bearing eddies with velocity $\vec{u}_0$ (\emph{i.e.} of wave-number $\vec k_0$). 
This time-scale is obtained by assuming uniform density, which is valid in the Sun for $\vec k > \vec k_0$ (\emph{i.e.} in the inertial sub-range) since the density scale-height  approximately equals the length-scale of energy-bearing eddies ($({\rm d}\ln \rho / {\rm d}r)^{-1} \approx 2\pi/k_0$). It also assumes the quasi-normal approximation \citep[see][ for details]{Kaneda93,Kaneda99, Rubinstein02}.  
The Eulerian micro-scale then corresponds to the timescale over which the energy-bearing eddies of velocity $\vec u_0$ advect eddies of size $2\pi k^{-1}$. It is the lowest time-scale (highest frequency) on which those eddies can be advected. 

\begin{figure}[t]
\begin{center}
\includegraphics[height=6cm,width=9cm]{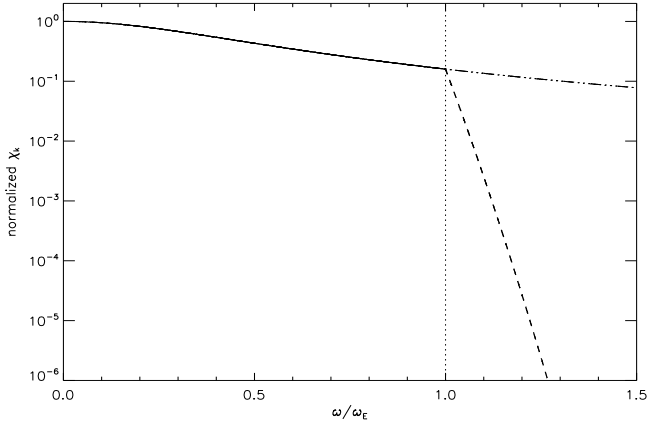}
\caption{Schematic time-correlation ($\chi_k$) versus normalized eddy frequency ($\omega/\omega_E$) at $k=5k_0$ (\emph{i.e.} in the inertial subrange such as $k_d \gg k > k_0$ ), where $k_0=6.28 \times 10^{-6} $m$^{-1}$. Note that the value of $k_0$ does not influence the result. 
The solid line (resp. dashed triple dot line) correponds to the Lorentizan functional form of $\chi_k$ for $\omega < \omega_E$ (resp. $\omega < \omega_E$). 
 The dashed line corresponds to a Gaussian modeling ($\chi_k \propto e^{-(\omega/\omega_E)^2}$) of characteristic frequency $\omega_E$. 
  (we numerically verified that $e^{-(\tau/\tau_E)^2}$ is a good approximation of \eq{DL}, see also Sect.~\ref{cut_off}). 
We stress that the sharp decrease the functional form given by \eq{DL}, in the temporal Fourier domain, then justifies to consider $\omega_E$ as a cut-off frequency. In other words, $\chi_k$ is computed according to \eq{chik_final}.}
\label{chi_k}
\end{center}
\end{figure}

\subsection{Eulerian time micro-scale as a cut-off frequency}
\label{cut_off}

The issue is now to estimate to what extent $\omega_E$ can be considered as a cut-off frequency, \emph{i.e.} that there is a sharp change in the slope of $\chi_k$ at high $\omega$. 

To this end, we first remark that the zero-th- and first-order terms in \eq{DL} are consistent with an exponential decrease of width $\alpha_k$ (\emph{i.e.} a Lorentzian in the frequency domain of width $\omega_k$, \eq{lorentzian}) for small $\tau$. In contrast, the zero-th-order term together with the second order term in \eq{DL} are consistent with a Gaussian behavior of width $\tau_E$. 
Hence, the relative importance of those two regimes depends on the relative magnitude of the second and third terms in \eq{DL}. 
Let us define the ratio ($\mathcal{R}$) of the first to the second order term in the expansion of $\mathcal{E}$ (\eq{DL}) 
\begin{align}
\label{ratioR}
\mathcal{R} = 2 \, \left( \omega_E \tau \right)^{-1} \;  \left(\frac{\omega_k}{\omega_E}\right) \, ,
\end{align} 
To evaluate this ratio, we compare the typical frequencies $\omega_k$ and $\omega_E$ using \eq{tau_E} together with \eq{omega_k}.  
Adopting a Kolmogorov spectrum $E(k) = C_K \, \epsilon^{2/3} \, k^{-5/3}$, with $C_K$ the Kolmogorov universal constant (close to 1.72), we have 
$u_k^2  =  \beta \, u_0^2 \, \left ( k / k_0 \right) ^{-2/3}$ , 
where $\beta=0.555$. 
Hence 
\begin{align}
\frac{\omega_E}{\omega_k} = \frac{u_0}{u_k} =  \beta^{-1/2} \,  \left(\frac{k}{k_0}\right)^{1/3} \, .
\label{tau_ratio_2}
\end{align}
From \eq{tau_ratio_2} we conclude that for $k \gg k_0$ (i.e. at small scale) we have $\omega_E / \omega_k \approx \beta^{-1/2} \, (k/k_0)^{1/3} \gg 1 $, then $\tau_E \ll \tau_k$. 
And for $k \approx k_0$ (i.e. at large scale), we have $\omega_E/\omega_k = \beta^{-1/2} \approx 1.4$. 
Hence, we always are in the situation where $\omega_E > \omega_k$. 

From \eq{ratioR}, it immediately follows that for $\omega \gtrsim \omega_E$ the second order term dominates over the first order one in \eq{DL}, at all  length-scales.  
We then conclude that for frequencies near the micro-scale frequency ($\omega \gtrsim \omega_E$), the eddy-time correlation function behaves as a Gaussian function $(e^{-(\omega/\omega_E)^2})$ instead of a Lorentzian function, resulting in a sharp decrease with $\omega$ (see Fig.~\ref{chi_k}). Hence, the contributions for $\omega > \omega_E$ are negligible and the temporal correlation is computed as follows
\begin{equation} 
\label{chik_final}
\chi_k(\omega) = \left\{ 
\begin{array}{rl} 
\frac{\displaystyle{1}}{\displaystyle{1+\left( \omega/\omega_k \right)^2}}  & \quad\text{if } \omega \le \omega_E\\ 
& \\
0  & \quad \text{if }  \omega > \omega_E \, . 
\end{array} \right. 
\end{equation} 

\section{Computation of the $p$-mode energy injection rates}
\label{amplitude}

\subsection{Computation of the energy injection rate}
\label{calcul_P}

The formalism we used to compute excitation rates of radial modes 
was developed by \cite{Samadi00I} and \cite{Samadi05c} \citep[see][ for a thorough discussion]{Samadi09}

For a radial mode of frequency $\omega_0=2\pi \, \nu_0$, the excitation rate (or equivalently, the energy injection rate), $P$,  mostly  arises from the 
Reynolds stresses  and can be written  as \citep[see Eq.~(21) of][]{Belkacem08} 
\begin{align}
\label{puissance}
P(\omega_0) &= \frac{\pi^{3}}{2  I} \int_{0}^{M} \, \left[ \rho_0  \, \left(\frac{16}{15}\right) \left(\frac{\partial \xi_r}{\partial r}\right)^2 \;  \int_{0}^{+\infty}  
 \mathcal{S}_k \; \textrm{d}k \right]  \textrm{d}m \\
\label{source}
\mathcal{S}_k &= \frac{E^2(k)}{k^2 } \int_{-\infty}^{+\infty}   
~\chi_k( \omega + \omega_0) ~\chi_k( \omega ) \; \textrm{d}\omega \, ,
\end{align}
where $\xi_r$ is the radial component of the fluid displacement eigenfunction ($\vec \xi$), $m$ is the local mass, $\rho_0$ the mean density, $\omega_0$ the mode angular 
frequency, $I$ the mode inertia,
$\mathcal{S}_k$ the source function,  $E(k)$ the spatial kinetic energy 
spectrum, $\chi_k$ the eddy-time correlation function, and $k$ the wave-number. 

The rate ($P$) at which energy is injected into a mode is
computed according to \eq{puissance}. 
In this letter, we consider two theoretical models, namely: 
\begin{itemize}
\item An analytical approach:  
the 1D calibrated solar structure model used for these computations is
obtained with the stellar evolution code CESAM2k  \citep{Morel97,Morel08}. 
The atmosphere is computed assuming an Eddington grey atmosphere.
Convection is included according to a B\"{o}hm-Vitense mixing-length (MLT) formalism
\citep[see][ for details]{Samadi06}, from which the convective velocity 
is computed. The mixing-length parameter $\alpha$ is adjusted in a way that the model reproduces the solar radius and the solar luminosity at the solar age. This calibration gives $\alpha= 1.90$, with an helium mass fraction of $0.245$, and a chemical composition following \cite{GN93}. The equilibrium model also includes turbulent pressure.  
\item A semi-analytical approach: calculation of the 
mode excitation rates is performed essentially in the manner of \cite{Samadi08,Samadi08b}. 
All required quantities, except the eddy-time correlation function, the mode eigenfunctions ($\xi_r$) and mode inertia ($I$), are directly obtained from 
a 3D simulation of the outer layers of the Sun \citep[see][ for details on the numerical simulation]{Samadi09}. 
\end{itemize}
In both cases, the eigenfrequencies and eigenfunctions  are computed
with the adiabatic pulsation code ADIPLS \citep{CD08}. We stress again that in both cases $\chi_k$ is implemented as an analytical function. 

\subsection{Results on mode amplitudes}
\label{results}

\begin{figure}[t]
\begin{center}
\includegraphics[height=6.5cm,width=9cm]{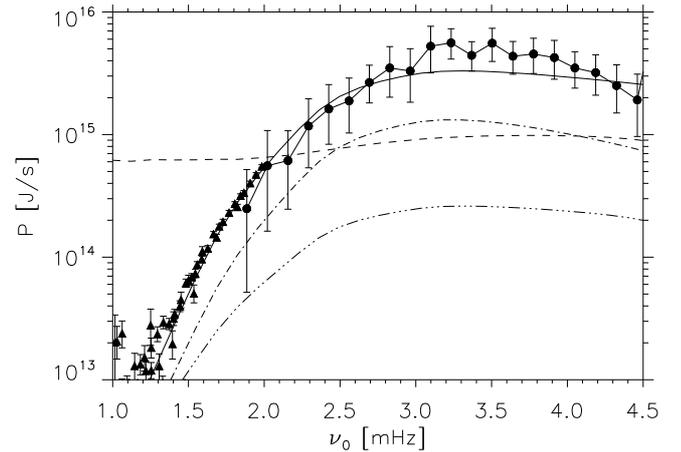}
\caption{
Solar $p$-mode excitation rates as a function of the frequency $\nu$. The dots correspond to the observational data obtained by the GONG network, as derived by \cite{Baudin05}, and the triangles corresponds to observational data obtained by the GONG network as derived by \cite{Salabert09} for $\ell=0$ to $\ell=35$. 
The dashed line corresponds to the computation of the excitation rates using the analytical approach together with a Lorentzien description of $\chi_k$ \emph{without} any cut-off frequency. Note that this modeling is similar to that mentioned by \cite{Chaplin05}. 
The solid line corresponds to the computation of mode excitation rates using the semi-analytical approach as described in Sect.~\ref{calcul_P} and using a Lorentzian $\chi_k$ together \emph{with} a cut-off frequency at $\omega=\omega_E$. 
The dashed triple dot line corresponds to the analytical approach using a Lorentzian description of $\chi_k$ down to the cut-off frequency $\omega_E$.  
Finally, the dashed-dot line corresponds to a semi-analytical approach using a Gaussian description of $\chi_k$. 
 Note that both solid (Lorentzian $\chi_k$) and dashed-dot (Gaussian $\chi_k$) lines present a similar frequency dependance, and since both are computed using the 3D numerical simulations for the convective motions the differences only comes from the way turbulence is temporally correlated.}
\label{resultat}
\end{center}
\end{figure}

When the frequency range of $\chi_k$ is extended toward infinity, computation of $P$ according to \eq{puissance} and \eq{source} fails to reproduce the observations, in particular the low-frequency shape. In order to illustrate this issue, we have computed the solar model excitation rates, using the solar global model described in Sect.~\ref{calcul_P}. The turbulent kinetic energy spectrum ($E(k)$) is assumed to be a Kolmogorov spectrum to be consistent with the derivation of $\tau_E$ as proposed by \cite{Kaneda93}. In addition, the eddy-time correlation function is supposed to be Lorentzian as described by \eq{lorentzian} for all $\omega >0$. 
In agreement with the results of \cite{Chaplin05} and \cite{Samadi09}, it results in an over-estimate of the excitation rates at low frequency (see Fig.~\ref{resultat} top). 
 
In contrast, by assuming that the time-dynamic of eddies in the Eulerian point of view is dominated by the sweeping, the Eulerian time micro-scale arises as a cut-off frequency (see Sect.~\ref{cut_off}). Hence, $\chi_k (\omega)$ is modeled following \eq{lorentzian} for $\omega < \omega_E$ 
and $\chi_k (\omega) = 0$ elsewhere. 
Using this procedure to model $\chi_k(\omega)$ (\emph{i.e.} by introducing $\omega_E$ as a cut-off frequency) permits us to reproduce the low-frequency ($\nu < 3$mHz) shape of the mode excitation rates as observed by the GONG network (see Fig.~\ref{resultat}).  
This is explained as follows: for large-scale eddies near $k_0^{-1}$, situated deep in the convective region, the cut-off frequency $\omega_E$  is close to $\omega_k$ as shown by \eq{tau_ratio_2}. As a consequence, the frequency range over which $\chi_k$ is integrated in \eq{source} is limited, resulting in lower injection rates into the modes. 
 
Note that the absolute values of mode excitation rates are not reproduced by using a mixing-length description of convection, this is in agreement with \cite{Samadi09}, and arises because that it underestimates the convective velocities as well as convective length-scales. It then explains the differences between the computation of mode excitation rates using the MLT and the 3D numerical simulations (Fig.~\ref{resultat}). 

\section{Conclusion and discussion}
\label{conclu}

By using a short-time analysis and the sweeping assumption, we have shown that there is a frequency ($\omega_E$ the micro-scale frequency) beyond the temporal correlation $\chi_k$ sharply decreases with frequency.
Including this frequency as a cut-off in our modeling of $\chi_k$ and assuming a Lorentzian shape we are able to reproduce the observed low-frequency ($\nu < 3$ mHz) excitation rates.

 These results then re-conciliate the theoretical and observational evidence that the frequency dependence of the eddy-time correlation may be Lorentzian in the whole solar convective region down to the cut-off frequency $\omega_E$.  
 Finally, it also represents a validation of the sweeping assumption in highly turbulent flows. 

We note, however, that one must remove several theoretical shortcomings to go further. 
For instance, a rigourous treatment of the energy-bearing eddies is needed.  
The short-time analysis and the computation of the Eulerian miscro-scale must be reconsidered  
by including the effect of  buoyancy that mainly affects large scales. 
Furthermore, some discrepancies remain at high-frequency ($\nu > 3$ mHz), and to go beyond these 
one has to remove the separation of scales assumption \citep[see][ for a dedicated discussion]{Belkacem08} and  include non-adiabatic effects. 

Eventually, we note that the modeling of amplitudes under the sweeping assumption is to be extended. In particular, the investigation of the effect of the sweeping assumption on solar gravity mode amplitudes is desirable.

\begin{acknowledgements} 
We are indebted to J. Leibacher for his careful reading of
the manuscript and his helpful remarks. 
K.B. acknowledges financial support 
through a postdoctoral fellowship from the 
ÒSubside f\'ed\'eral pour la recherche 2010Ó, University of Li\`ege. 
The authors also acknowledge financial support from the 
French National Research Agency (ANR) for the SIROCO (SeIsmology, 
ROtation and COnvection with the COROT satellite) project.
\end{acknowledgements}


\begin{thebibliography}{37}
\expandafter\ifx\csname natexlab\endcsname\relax\def\natexlab#1{#1}\fi

\bibitem[{{Appourchaux} {et~al.}(2010){Appourchaux}, {Belkacem}, {Broomhall},
  {Chaplin}, {Gough}, {Houdek}, {Provost}, {Baudin}, {Boumier}, {Elsworth},
  {Garc{\'{\i}}a}, {Andersen}, {Finsterle}, {Fr{\"o}hlich}, {Gabriel}, {Grec},
  {Jim{\'e}nez}, {Kosovichev}, {Sekii}, {Toutain}, \&
  {Turck-Chi{\`e}ze}}]{Appourchaux10}
{Appourchaux}, T., {Belkacem}, K., {Broomhall}, A., {et~al.} 2010, \aapr, 18,
  197

\bibitem[{{Balmforth}(1992)}]{B92c}
{Balmforth}, N.~J. 1992, \mnras, 255, 639

\bibitem[{{Baudin} {et~al.}(2005){Baudin}, {Samadi}, {Goupil}, {Appourchaux},
  {Barban}, {Boumier}, {Chaplin}, \& {Gouttebroze}}]{Baudin05}
{Baudin}, F., {Samadi}, R., {Goupil}, M.-J., {et~al.} 2005, \aap, 433, 349

\bibitem[{{Belkacem} {et~al.}(2010){Belkacem}, {Dupret}, \&
  {Noels}}]{Belkacem10b}
{Belkacem}, K., {Dupret}, M.~A., \& {Noels}, A. 2010, \aap, 510, A6+

\bibitem[{{Belkacem} {et~al.}(2009{\natexlab{a}}){Belkacem}, {Samadi},
  {Goupil}, {Lef{\`e}vre}, {Baudin}, {Deheuvels}, {Dupret}, {Appourchaux},
  {Scuflaire}, {Auvergne}, {Catala}, {Michel}, {Miglio}, {Montalban}, {Thoul},
  {Talon}, {Baglin}, \& {Noels}}]{Belkacem09b}
{Belkacem}, K., {Samadi}, R., {Goupil}, M., {et~al.} 2009{\natexlab{a}},
  Science, 324, 1540

\bibitem[{{Belkacem} {et~al.}(2008){Belkacem}, {Samadi}, {Goupil}, \&
  {Dupret}}]{Belkacem08}
{Belkacem}, K., {Samadi}, R., {Goupil}, M.-J., \& {Dupret}, M.-A. 2008, \aap,
  478, 163

\bibitem[{{Belkacem} {et~al.}(2009{\natexlab{b}}){Belkacem}, {Samadi},
  {Goupil}, {Dupret}, {Brun}, \& {Baudin}}]{Belkacem09}
{Belkacem}, K., {Samadi}, R., {Goupil}, M.~J., {et~al.} 2009{\natexlab{b}},
  \aap, 494, 191

\bibitem[{{Chaplin} {et~al.}(2005){Chaplin}, {Houdek}, {Elsworth}, {Gough},
  {Isaak}, \& {New}}]{Chaplin05}
{Chaplin}, W.~J., {Houdek}, G., {Elsworth}, Y., {et~al.} 2005, \mnras, 360, 859

\bibitem[{{Christensen-Dalsgaard}(2008)}]{CD08}
{Christensen-Dalsgaard}, J. 2008, \apss, 316, 113

\bibitem[{{Dolginov} \& {Muslimov}(1984)}]{Dolginov84}
{Dolginov}, A.~Z. \& {Muslimov}, A.~G. 1984, \apss, 98, 15

\bibitem[{{Dupret} {et~al.}(2009){Dupret}, {Belkacem}, {Samadi}, {Montalban},
  {Moreira}, {Miglio}, {Godart}, {Ventura}, {Ludwig}, {Grigahc{\`e}ne},
  {Goupil}, {Noels}, \& {Caffau}}]{Dupret09}
{Dupret}, M., {Belkacem}, K., {Samadi}, R., {et~al.} 2009, \aap, 506, 57

\bibitem[{{Goldreich} \& {Keeley}(1977)}]{GK77}
{Goldreich}, P. \& {Keeley}, D.~A. 1977, \apj, 212, 243

\bibitem[{{Goldreich} {et~al.}(1994){Goldreich}, {Murray}, \& {Kumar}}]{GK94}
{Goldreich}, P., {Murray}, N., \& {Kumar}, P. 1994, \apj, 424, 466

\bibitem[{{Gough}(1977)}]{Gough77}
{Gough}, D.~O. 1977, \apj, 214, 196

\bibitem[{{Grevesse} \& {Noels}(1993)}]{GN93}
{Grevesse}, N. \& {Noels}, A. 1993, in Origin and Evolution of the Elements,
  ed. {N.~Prantzos, E.~Vangioni-Flam, \& M.~Casse}, 15--25

\bibitem[{{Houdek}(2009)}]{Houdek09}
{Houdek}, G. 2009, \apss, 252

\bibitem[{{Kaneda}(1993)}]{Kaneda93}
{Kaneda}, Y. 1993, Physics of Fluids, 5, 2835

\bibitem[{{Kaneda} {et~al.}(1999){Kaneda}, {Ishihara}, \& {Gotoh}}]{Kaneda99}
{Kaneda}, Y., {Ishihara}, T., \& {Gotoh}, K. 1999, Physics of Fluids, 11, 2154

\bibitem[{{Morel}(1997)}]{Morel97}
{Morel}, P. 1997, \aaps, 124, 597

\bibitem[{{Morel} \& {Lebreton}(2008)}]{Morel08}
{Morel}, P. \& {Lebreton}, Y. 2008, \apss, 316, 61

\bibitem[{{Rubinstein} \& {Zhou}(2002)}]{Rubinstein02}
{Rubinstein}, R. \& {Zhou}, Y. 2002, \apj, 572, 674

\bibitem[{{Salabert} {et~al.}(2009){Salabert}, {Leibacher}, {Appourchaux}, \&
  {Hill}}]{Salabert09}
{Salabert}, D., {Leibacher}, J., {Appourchaux}, T., \& {Hill}, F. 2009, \apj,
  696, 653

\bibitem[{{Samadi}(2009)}]{Samadi09}
{Samadi}, R. 2009, Stochastic excitation of acoustic modes in stars, ArXiv
  e-prints (0912.0817)

\bibitem[{{Samadi} {et~al.}(2008{\natexlab{a}}){Samadi}, {Belkacem}, {Goupil},
  {Ludwig}, \& {Dupret}}]{Samadi08}
{Samadi}, R., {Belkacem}, K., {Goupil}, M., {Ludwig}, H., \& {Dupret}, M.
  2008{\natexlab{a}}, Communications in Asteroseismology, 157, 130

\bibitem[{{Samadi} {et~al.}(2008{\natexlab{b}}){Samadi}, {Belkacem}, {Goupil},
  {Dupret}, \& {Kupka}}]{Samadi08b}
{Samadi}, R., {Belkacem}, K., {Goupil}, M.~J., {Dupret}, M., \& {Kupka}, F.
  2008{\natexlab{b}}, \aap, 489, 291

\bibitem[{{Samadi} {et~al.}(2007){Samadi}, {Georgobiani}, {Trampedach},
  {Goupil}, {Stein}, \& {Nordlund}}]{Samadi07}
{Samadi}, R., {Georgobiani}, D., {Trampedach}, R., {et~al.} 2007, \aap, 463,
  297

\bibitem[{{Samadi} {et~al.}(2001){Samadi}, {Goupil}, \&
  {Lebreton}}]{Samadi00II}
{Samadi}, R., {Goupil}, M.~., \& {Lebreton}, Y. 2001, \aap, 370, 147

\bibitem[{{Samadi} \& {Goupil}(2001)}]{Samadi00I}
{Samadi}, R. \& {Goupil}, M.~J. 2001, \aap, 370, 136

\bibitem[{{Samadi} {et~al.}(2005){Samadi}, {Goupil}, {Alecian}, {Baudin},
  {Georgobiani}, {Trampedach}, {Stein}, \& {Nordlund}}]{Samadi05c}
{Samadi}, R., {Goupil}, M.-J., {Alecian}, E., {et~al.} 2005, J. Astrophys.
  Atr., 26, 171

\bibitem[{{Samadi} {et~al.}(2006){Samadi}, {Kupka}, {Goupil}, {Lebreton}, \&
  {van't Veer-Menneret}}]{Samadi06}
{Samadi}, R., {Kupka}, F., {Goupil}, M.~J., {Lebreton}, Y., \& {van't
  Veer-Menneret}, C. 2006, \aap, 445, 233

\bibitem[{{Samadi} {et~al.}(2010{\natexlab{a}}){Samadi}, {Ludwig}, {Belkacem},
  {Goupil}, {Benomar}, {Mosser}, {Dupret}, {Baudin}, {Appourchaux}, \&
  {Michel}}]{Samadi10b}
{Samadi}, R., {Ludwig}, H., {Belkacem}, K., {et~al.} 2010{\natexlab{a}}, \aap,
  509, A16+

\bibitem[{{Samadi} {et~al.}(2010{\natexlab{b}}){Samadi}, {Ludwig}, {Belkacem},
  {Goupil}, \& {Dupret}}]{Samadi10a}
{Samadi}, R., {Ludwig}, H., {Belkacem}, K., {Goupil}, M.~J., \& {Dupret}, M.
  2010{\natexlab{b}}, \aap, 509, A15+

\bibitem[{{Samadi} {et~al.}(2003{\natexlab{a}}){Samadi}, {Nordlund}, {Stein},
  {Goupil}, \& {Roxburgh}}]{Samadi02II}
{Samadi}, R., {Nordlund}, {\AA}., {Stein}, R.~F., {Goupil}, M.~J., \&
  {Roxburgh}, I. 2003{\natexlab{a}}, \aap, 404, 1129

\bibitem[{{Samadi} {et~al.}(2003{\natexlab{b}}){Samadi}, {Nordlund}, {Stein},
  {Goupil}, \& {Roxburgh}}]{Samadi02I}
{Samadi}, R., {Nordlund}, {\AA}., {Stein}, R.~F., {Goupil}, M.~J., \&
  {Roxburgh}, I. 2003{\natexlab{b}}, \aap, 403, 303

\bibitem[{{Stein}(1967)}]{Stein67}
{Stein}, R.~F. 1967, Solar Physics, 2, 385

\bibitem[{{Tennekes}(1975)}]{Tennekes75}
{Tennekes}, H. 1975, Journal of Fluids Mechanics, 67, 561

\bibitem[{{Tennekes} \& {Lumley}(1972)}]{Tennekes_Lumley72}
{Tennekes}, H. \& {Lumley}, J.~L. 1972, {First Course in Turbulence}, ed.
  {Tennekes, H.~\& Lumley, J.~L.}

\end{thebibliography}
\end{document}